\documentclass[aps,prl,showpacs]{revtex4}
\usepackage{graphicx}
\newcommand{\omu}{{\overline\mu}}

\newcommand{\fN}{{\frac{1}{N}}}

\begin{document}
\title{Mixed phase and bound states in the phase diagram of the extended Hubbard model}
\author{Maciej Bak}
\email[]{karen@delta.amu.edu.pl} \affiliation{Institute of
Physics, A. Mickiewicz University, Umultowska 85, 61-614 Pozna\'n,
Poland}

\begin{abstract}
 The paper examines part of the
ground state diagram of the extended Hubbard model, with the
on-site attraction $U<0$ and intersite repulsion $W>0$ in the
presence of charge density waves, superconducting and
$\eta$-superconducting order parameters. We show the possibility
of the stabilization of the mixed state, with all three nonzero
order parameters, in the model with nearest neighbor interactions.
The other result of the paper is application of the exact solution
of the Schrodinger equation for the two-electron bound state, as
an additional bound for the phase diagram of the model, resulting
in the partial suppression of the superconducting state of the
s-wave symmetry, in favor of the normal state phase.
\end{abstract}


\maketitle

\section{Introduction}
The Hubbard model is known for its universality. The model and its
extensions have been used in research of magnetism,
superconductivity and especially high temperature
superconductivity (HTS), other various phases in the solid, charge
orderings, phase separation etc., in materials like high
temperature superconductors, bismuthates, Chevrel phases,
amorphous semiconductors, heavy fermion materials, systems with
alternating valence to name the few (for a review see, e.g.,
Ref.\cite{review}). Despite intensive research the results
concerning the model come mostly from different approximations or
numerical simulations. The exact results are scarce and concern
mostly specific situation of one dimension (1D) or specific band
fillings or specific limits: zero bandwidth, infinite $U$ or
infinite $d$. We can mention the exact solution in 1D obtained by
Bethe ansatz\cite{betheA}, Nagaoka ferromagnetism in repulsive,
half-filled systems with one hole\cite{nagaoka}, flat band
ferromagnetism\cite{mielke}, Lieb's ferrimagnetism\cite{lieb},
some bounds on correlation functions\cite{koma,kubo,shen} and a
statement of necessity of bound states for existence of extended
s-wave superconductivity in 2D systems with low electron density
$n$, by Randeria\cite{randeria}. Let's also note that the
mean-field BCS equations for superconductivity in the Hubbard
model with effective attractive interaction between electrons, in
the limit of vanishing electron density turn into Schrodinger
equations, which can also be solved exactly\cite{nozieres}.

An interesting, unresolved question concerning Hubbard model is:
what happens, when we increase the strength of the
 interaction? What phases do we obtain? Or does the system prefer
 phase separation? That was the conclusion of the paper by
 Robaszkiewicz and Pawlowski\cite{rp}. By the broken symmetry
 Hartree-Fock approach they established that the attractive extended Hubbard
 model (with nearest neighbor -- nn -- density-density interactions of strength
 $W$ added) in two dimensions for arbitrary electron density and moderate values
  of $|W|$ is superconducting in the ground state but
 becomes phase separated with increasing $|W|$ (except
 from $n=1$, where the ground state is in charge density wave phase for any $W\ge 0$).
 In the case of $W<0$ they obtained the phase of electron droplets and for $W>0$
 a phase separation between charge density waves (CDW) with $n=1$
 and a singlet on-site superconductivity (SS) (extended s-wave was not considered).
 The free energy of this phase separated state  (PS[CDW/SS]) turned out to
 be lower than the energy of the homogenous mixed state
 (M(CDW+SS)). Stabilization of M state required longer than nn
repulsive intersite interactions ($W_2$) or increasing $|U|$. For
$U=-\infty$ PS and M states were degenerated even for $W_2=0$.

 In this paper it
 will be shown that including $\eta$-pairing (superconducting pairing
 with singlet pairs with center-of-mass momentum $\pi$)\cite{yang,penson,robbulka}
 into the
 Hamiltonian stabilizes mixed phase, already with intersite
 interactions $W$ restricted to the nn (in disagreement with
 Ref.\cite{rp}).
 It turns out that the phase separation PS[CDW/SS] exists only below certain critical
value of $W$, above which the M phase can be stabilized.

\section{Low density limit}
It can be shown, that in the low density limit the BCS equations
go over to the Schrodinger equation for the bound
pair\cite{nozieres}. In the center-of mass coordinate system the
wave function of the two particle bound state can be expanded  as
the sum of states with the definite relative momenta. Using that
expansion we can obtain a set of self-consistent equations for the
pair wave function in a position space, in terms of lattice Green
functions\cite{review,blaer}. For the hypercubic lattices these
equations were solved (see, e.g., \cite{review}) and it has been
found out that in one and two dimensions pairs for $W=0$ bind for
any negative $U$, while in three dimensions there is critical
value for $W$. In the cited reference the formula was given for
the $W_{cr}$ as a function of $U$:
\begin{equation}\label{wcrex}
    \frac{|W_{cr}|}{2t}=\left[1+\frac{2D}{U}\right]^{-1}+(C-1)^{-1}
\end{equation}
where $C=1/N\sum_k(1-\gamma_k/z)^{-1}$ is the Watson integral. In
 Ref.\cite{review} this formula was given only for the case of
$W<0$ but in fact it is valid for any  $U$ and $W$. What's more,
there exists analogic formula, applicable not only in the low
density limit but for any electron density. It was derived from
the Hamiltonian after Hartree-Fock linearization so it is not
exact, but goes over into the exact Eq.~(\ref{wcrex}) for
$n\rightarrow 0$. The formula is given by:
\begin{equation}\label{wcrmy}
    W_{cr}=\frac{8t^2}{\mu(1-n)+8tI-2\mu^2/U}
        \mbox{,\hspace{0.5cm}where\hspace{0.5cm}}I=\int_{-1}^{\mu}x\rho(x)\;dx
\end{equation}
or for the rectangular DOS\cite{bak2sol}:
$W_{cr}(1+(n-1)^2(1+16t/U))=-4t$.

\section{Phase separation}
The energy of a phase separated state, $E_{PS}$, is calculated
according to the formula:
\begin{equation}\label{Eps}
    E_{PS}=mE_{+}(n_{+})+(1-m)E_{-}(n_{-})
\end{equation}
where $E_\pm(n_\pm)$ are the values of $<H>/N$ of the two
separating phases at the $n=n_\pm$ corresponding to the lowest
energy homogenous solution for a given phase. $m$ and $(1-m)$ are
 fractions of the system occupied by the phases $E_+$ and $E_-$
respectively. There is also a similar equation connecting charge
densities in the two coexisting phases:
\begin{equation}\label{MaxFrac}
    mn_{+}+(1-m)n_{-}=n
\end{equation}
A very useful is also Maxwell construction:
\begin{equation}\label{Max}
    \frac{\partial E_{-}(n_{-})}{\partial n_{-}}=\frac{E_{+}(n_{+})-E_{-}(n_{-})}{n_{+}-n_{-}}
\end{equation}
which is equivalent to the condition of minimization of $E_{PS}$
respectively $n_{-}$ with fixed $n_+$ and $n$. Taking into account
Eqs~(\ref{MaxFrac}) and (\ref{Max}), Eq.~(\ref{Eps}) can be
rewritten as:
\begin{equation}\label{Eps2}
    E_{PS}=E_{+}(n_{+})-(n_{+}-n)\frac{\partial E_{-}(n_{-})}{\partial n_{-}}
\end{equation}

There is also another, equivalent but more intuitive approach using
Gibbs energies $G=F-\mu n$. The independent variable for $G$ is
chemical potential $\mu$ instead of charge density $n$ for $F$.
Let's note, that instead of the analogue of  Eq.~(\ref{MaxFrac}),
with $\mu_\pm$ instead of $n_\pm$, we have $\mu_{+}=\mu_{-}$,
because this is the condition of equilibrium, or rather the
definition of equilibrium! (Eq.~(\ref{MaxFrac}) does not apply also
because $\mu$ is not an extensive quantity). There is only one value
of chemical potential, fixed and the same for both phases. This way
the whole construction simplifies. The procedure is as follows: we
find solutions for the homogenous phases, then we plot their Gibbs
energies vs $\mu$ in one figure and we look for the intersections of
$G$'s belonging to the different phases. Such intersection means a
coexistence of the two phases with the same $\mu$, i.e., in
equilibrium. The same value of $\mu$ usually translates into {\em
different} values of $n_{+}$ and $n_{-}$ in the two phases under
consideration and that's nothing else but the (definition of) phase
separation, which appears on the $F$-$n$ phase diagram for the
electron density $n$, constructed according to the
Eq.~(\ref{MaxFrac}).

Some remarks are due:
\begin{description}
    \item[--] let's note that the use of $G$ and chemical
    potential makes it clear that $n_{+}$ and $n_{-}$ are
    independent of $n$, as there is usually one intersection point
    of Gibbs energies.
    In other words, $m$ is a linear function of $n$:
    \begin{equation}\label{mfunc}
        m=\frac{1}{n_{+}-n_{-}}\;n-\frac{n_{-}}{n_{+}-n_{-}}
    \end{equation}

    \item[--] let's note that Eq.~(\ref{Eps}) gives boundaries for
    $E_{PS}$. By the definition in Eq.~(\ref{Eps}), $E_{PS}$ must be somewhere between
    $E_{+}(n_{+})$ and $E_{-}(n_{-})$. By the same reasoning that led us to the
    Eq.~(\ref{mfunc}) we can say that the dependence of phase
    separated states on $n$ should be strictly linear, with the
    same coefficients as in the Eq.~(\ref{mfunc}). That's nothing
    else but the Maxwell condition restated

        \item[--] the same considerations as above apply to the
    electron density: phase separated state can exist only for charge densities
    $n_{-}<n<n_{+}$. This is an important bound, which can be
    easily overlooked, as there are formal solutions for free energy of
    phase separated state also outside that range

    \item[--] a simple implication of the preceding points, which we will
    call {\em Theorem 1}: we can not obtain a phase
    transition by changing $n$ between two phase
    separated states having "a state in common" for the same fixed
    density $n_{+}$, i.e. we can not have phase transition between
    states like PS1[CDW(n=1)/phase1] and PS2[CDW(n=1)/phase2]. The
    lines depicting $F_{PS1}$ and $F_{PS2}$ will begin in the same
    point and will extend in different directions.

    \item[--]and last but not the least is the notion, that we have
    also bounds for homogenous phases: the phases can exist only for the
    positive compressibility, i.e. in the
    range of $\mu$, for which $\partial\mu/\partial n>0$
\end{description}

\section{The model} We start from the
extended Hubbard model:
\begin{equation}\label{ham}
    H=\sum_{ij\sigma}t_{ij}c^\dagger_{i\sigma}c_{j\sigma}+
        U \sum_i n_{i\uparrow}n_{i\downarrow}+
        \frac{1}{2}W \sum_{ij}n_{i\sigma}n_{j\sigma'}-
        \mu \sum_i n_i
\end{equation}
in standard notation. We use  Hartree-Fock approach. In the
mean-field Hamiltonian we retain charge density order parameter
$\Delta_c$ and singlet superconductivity order parameters:
$\Delta_0$ and $\Delta_Q$, for on-site and $\eta$-pairing
respectively (see also Ref.\cite{rp}):
\begin{eqnarray}\label{H0}\nonumber
    H_0&=&\sum_{k\sigma}(\varepsilon_k-\overline\mu)c^\dagger_{k\sigma}c_{k\sigma}+
        \frac{1}{2}\sum_{k\sigma}(\Delta_c
        c^\dagger_{k\sigma}c_{k+Q\sigma}+h.c.)+\\
        &&\sum_k (\Delta_0c^\dagger_{k\uparrow}c^\dagger_{-k\downarrow}+h.c)+
        \sum_k (\Delta_Q c^\dagger_{k+Q\uparrow}c^\dagger_{-k\downarrow}+h.c)+
        C
\end{eqnarray}
where the order parameters are given by:
\begin{eqnarray}
    \Delta_c&=&\frac{1}{N}\sum_{k\sigma}(-zW+\frac{U}{2})<c^\dagger_{k+Q\sigma}c_{k\sigma}>\\
    \Delta_0&=&\frac{U}{N}\sum_{k_2}<c_{-k_2\downarrow}c_{k_2\uparrow}>\\
    \Delta_Q&=&\frac{U}{N}\sum_{k_2}<c_{k_2\downarrow}c_{-k_2+Q\uparrow}>
\end{eqnarray}
where $z$ is number of nearest neighbors,
$\overline\mu=\mu-(\frac{U}{2}+z W)n$, $W_k=W\gamma_k$,
$\varepsilon_k=-t\gamma_k$, $\gamma_k=2\sum_{\alpha}^d\cos
k_\alpha$, $\alpha\in(x,y,z)$ and $d=z/2$ is dimensionality of the
hypercubic lattice;
$\varepsilon_k\in<-\varepsilon_0,\varepsilon_0>$ where
$\varepsilon_0\equiv D=zt$. In the following all values of
dimension energy will be given in units of $D$ and the simplest,
zero temperature forms of the equations will be most often used.
The constant $C$ takes the form:
\begin{equation}
    C=\frac{1}{4}(U+2W z)n^2-\frac{|\Delta_c|^2}{U-2Wz}-\frac{|\Delta_0|^2}{U}-\frac{|\Delta_Q|^2}{U}
\end{equation}
After diagonalization we obtain Hamiltonian $H_0$ with two
eigenenergies $A^\pm_k$:
\begin{equation}\label{apm}
    A^\pm_k=\sqrt{\varepsilon_k^2+\overline\mu^2+\Delta_0^2+\Delta_Q^2+\Delta_c^2\pm
        2\sqrt{\varepsilon_k^2(\overline\mu^2+\Delta_Q^2)+(\overline\mu\Delta_c-\Delta_0\Delta_Q)^2}}
\end{equation}
In terms of the eigenstates of $H_0$ the free energy is given by
the standard formula:
\begin{equation}
    F_0=-\frac{1}{\beta}\ln Tr\exp(-\beta H_0)+\mu N_e
\end{equation}
what yields:
\begin{equation}
    \frac{F_0}{N}=\overline\mu(n-1)+C-
    \frac{1}{\beta N}\sum_k\ln(4\cosh\frac{\beta A_k^+}{2}\cosh\frac{\beta A_k^-}{2})
\end{equation}
The equations for the order parameters and chemical potential are
obtained in a standard way, by minimization of the above
expression. For the most general mixed, homogenous phase with all
the order parameters present the equations take the form:
\begin{eqnarray}\label{m1}
    2(n-1)&=&\omu(I_0+J_2)+\Delta_c(\overline\mu\Delta_c-\Delta_0\Delta_Q)J_0\\
    \frac{-4\Delta_Q}{U}&=&\Delta_Q(I_0+J_2)-\Delta_0(\overline\mu\Delta_c-\Delta_0\Delta_Q)J_0\\
    \frac{-4\Delta_0}{U}&=&\Delta_0I_0-\Delta_Q(\overline\mu\Delta_c-\Delta_0\Delta_Q)J_0\\\label{m4}
    \frac{-4\Delta_c}{U-2Wz}&=&\Delta_cI_0+\omu\,(\overline\mu\Delta_c-\Delta_0\Delta_Q)J_0
\end{eqnarray}
where
\begin{eqnarray}\label{i0}
    I_0&=&\fN\sum_k\left(\frac{1}{A_k^+}+\frac{1}{A_k^-}\right)\\\label{jn}
    J_n&=&\fN\sum_k\left(\frac{1}{A_k^+}-\frac{1}{A_k^-}\right)\frac{\varepsilon_k^n}
    {\sqrt{\varepsilon_k^2(\overline\mu^2+\Delta_Q^2)+(\overline\mu\Delta_c-\Delta_0\Delta_Q)^2}}
\end{eqnarray}
where $A_k^\pm$ are given by Eq. (\ref{apm}). The equations for
the mixed phases with smaller number of constituent phases (i.e.,
M(CDW+SS), M(CDW+SQ), M(SS+SQ)) as well as for pure phases can be
obtained from Eqs. (\ref{m1})-(\ref{jn}) by just dropping out
respective order parameters. For the detailed form of the pure
phase equations the reader is referred, e.g. to Ref.\cite{rp,rmc}.

\section{Results}
The results were obtained  using rectangular density of states
(DOS) and $z=4$ as an approximation for two dimensional square
lattice. In Figs~1 (a) and (b) we plotted the results for the
thermodynamical potentials. In (a) we have shown all the solutions
for Gibbs energy (vs chemical potential), without applying yet the
bonds for $n\pm$ or checking the sign of $\partial\mu/\partial n$.
All homogenous phases are shown, except from $\eta$-pairing
(denoted by SQ) and mixed M(SS+SQ) (SS denoting singlet, on-site
superconductivity), which do not have solutions for $U>-2$.
Nevertheless the phases: M(CDW+SQ), M(CDW+SS), upper branch of
M(CDW+SS+SQ) (see lower inset) and upper branch of CDW for $n\neq
1$ must be excluded, because of the negative compressibility,
connected with the downward curvature of these lines (compare also
the upper inset - decreasing and increasing part of the plot
corresponds to the opposite signs of compressibility and to the
two branches of M(CDW+SS+SQ), from now on denoted as M3, in the
main figure).

\begin{figure}
    \includegraphics[width=8.5cm]{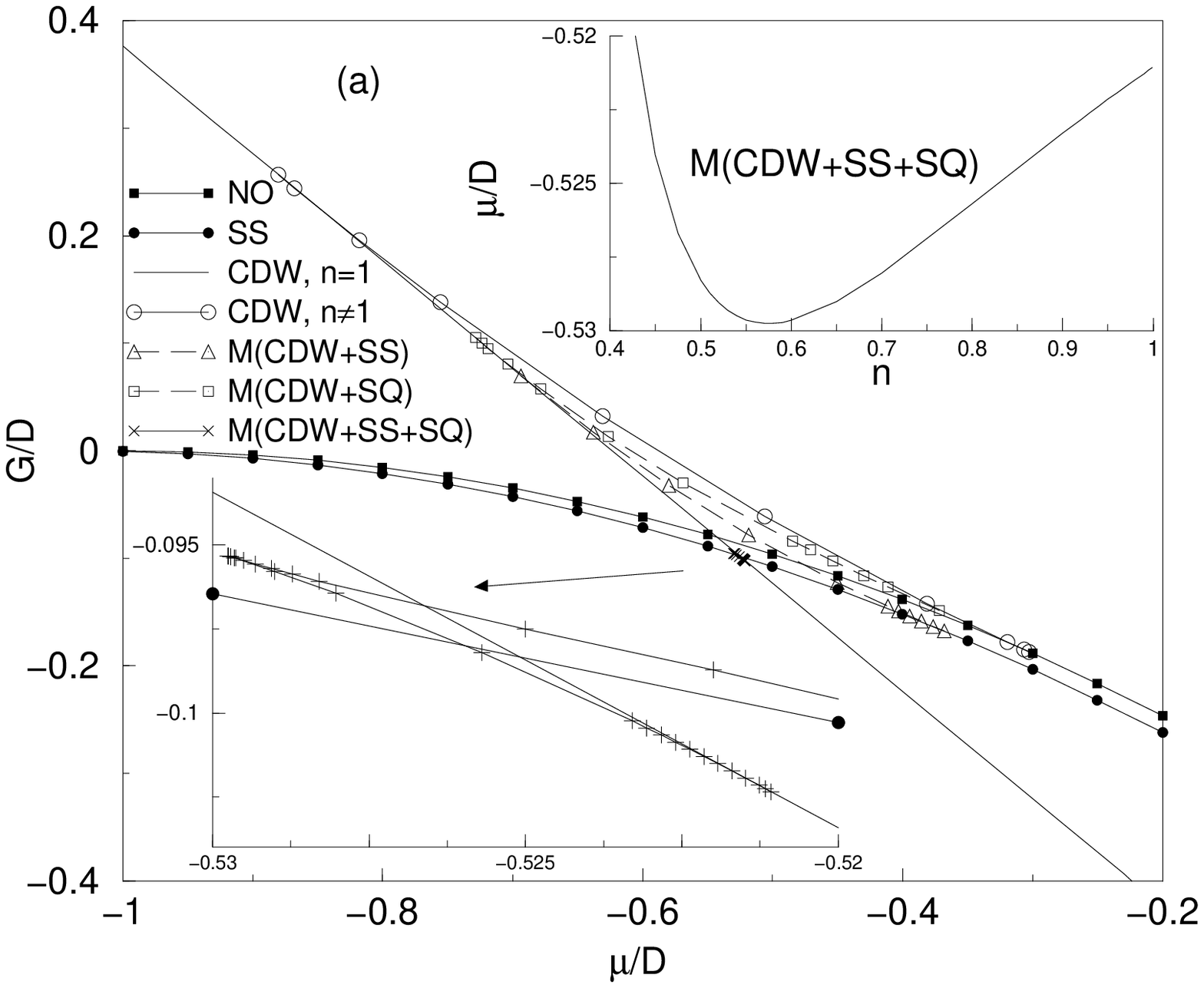}
    \includegraphics[width=8.5cm]{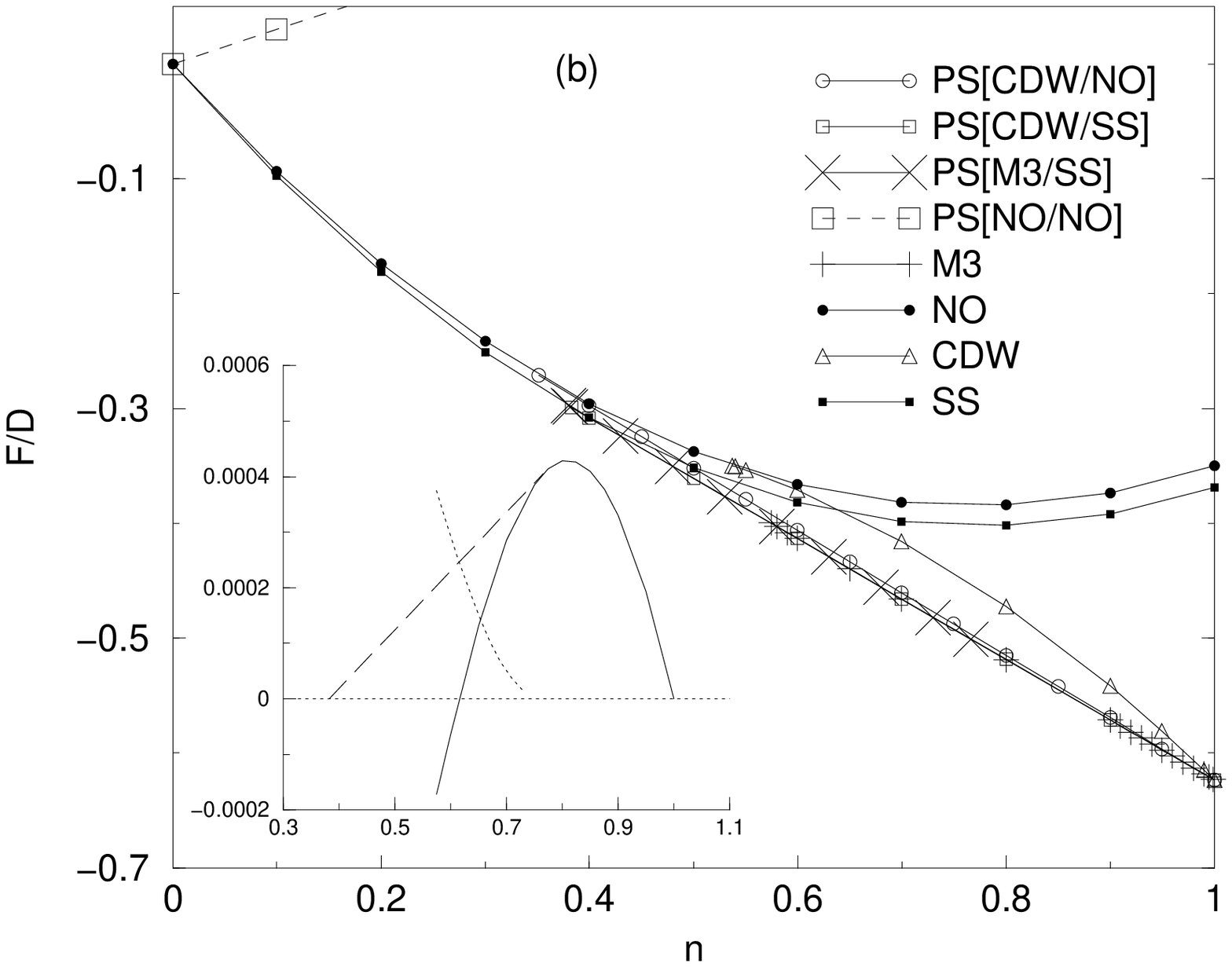}
    \caption{\label{wnPhDg} (a) Gibbs energies $G$ vs bare chemical
    potential $\mu$, for
    $U=-1$ and $W=0.2$. Lines are described in the figure. In the lower, left-hand corner
    in the inset, there is an enlargement of the
    area where the curves cross. In the inset in the upper right-hand corner there is plotted
    the dependence of chemical potential $\mu$ on charge density $n$ in M3 phase.
    (b) The free energy $F$ vs electron density
    $n$ for $U=-1$ and $W=0.2$. Lines are described in the figure. In the inset
    there are the differences of free energies plotted -- full line: $F(PS[CDW/SS])-F(M3)$,
    dashed line: $F(PS[CDW/SS])-F(PS[M3/SS])$, dotted line: $F(M3)-F(PS[M3/SS])$
    for the range of $n$, where $\partial\mu/\partial n>0$ for each phase. Horizontal
    dotted line is the just a guide for an eye. }
\end{figure}

The remaining curves (and CDW one, for comparison) are a basis for
plotting the free energy $F$-$n$ diagram, shown in (b). The plot
contains curves in the proper range of $n$, with all the bounds
already applied. Apart from all the homogenous phases fulfilling
the condition $\partial\mu/\partial n>0$ (what corresponds to the
downward curvature for $F(n)$) we can see also phase separated
states indicated in (a) by Gibbs energy plots crossings:
PS[CDW/SS], PS[CDW/NO], PS[M3/SS] (and PS[NO/NO], of slightly
different nature). The states phase separated with CDW(n=1) are
connecting the point of energy $E^{CDW}(n=1)$ with respective
curves of the other state coming into separation, in a tangential
way, in accordance with Maxwell construction and our earlier
remarks. Let's note, that the CDW inside PS states is a phase with
$n=1$ while CDW in mixed states is for $n\neq 1$. The states of
the lowest energy are: SS, M3, PS[CDW/SS] and PS[M3/SS] and it
will be so also for other values of parameters. In the inset there
are the energy differences between the states: PS[CDW/SS],
PS[M3/SS] and M3, difficult to discern in the main figure. We can
see that the phase separated state PS[M3/SS] has the lowest energy
of all the three (if exists). Its appearance is an unavoidable
consequence of {\em Theorem 1} in the case when the phase M3
starts to win with PS[CDW/SS]. The phase diagrams are shown in
Fig.~2~(a) and (b).

\begin{figure}[b]
    \includegraphics[width=8.5cm]{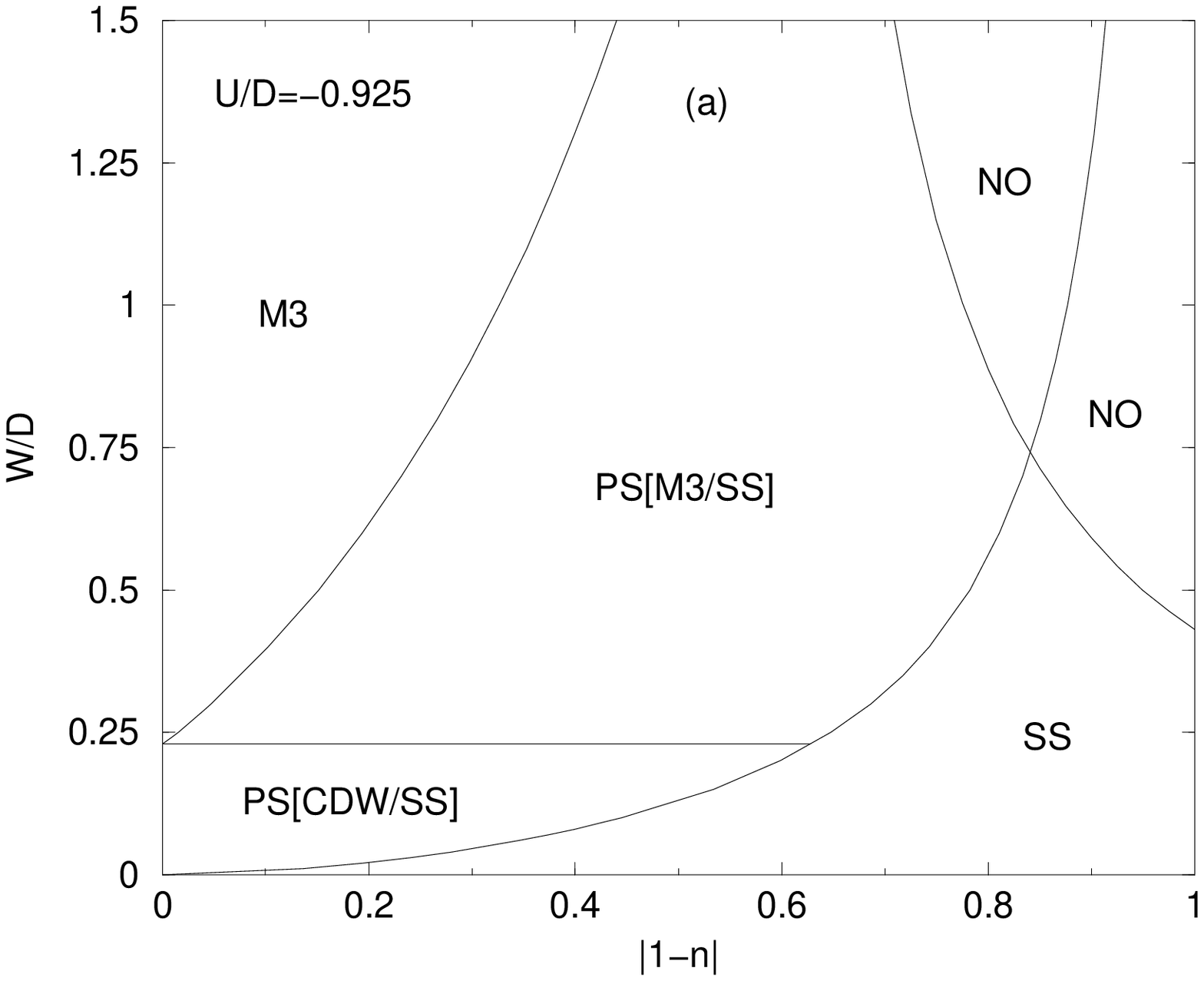}
    \includegraphics[width=8.5cm]{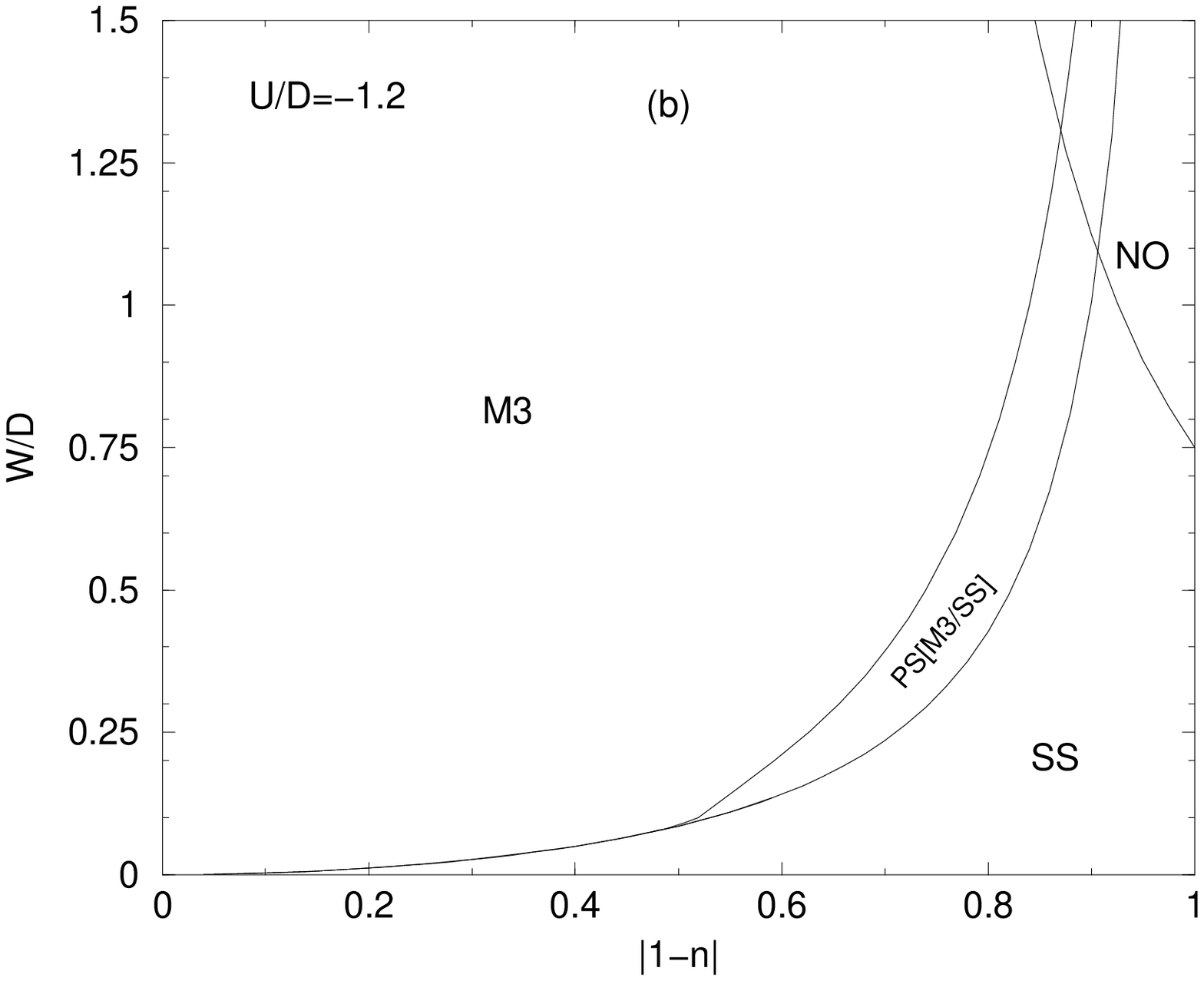}
    \caption{\label{wnPhDg} Ground state phase diagram of model
    Eq.~(\ref{H0}) calculated for the rectangular density of states for (a)
    $U=-0.925$  and  (b) $U=-1.2$. For the $n=1$ the ground state is
    CDW phase, a smooth limit of both M(CDW+SS+SQ) and PS[CDW/SS]
    phases.}
\end{figure}

For not too large $|U|$ and $W$ the ground state is in PS[CDW/SS]
or SS state, depending on $n$, in accordance with Ref.\cite{rp}.
When we increase $W$ we reach a threshold, smallest for the
half-filled band, above which the M3 phase is stabilized.
Appearing of the M3 phase is connected with replacing PS[CDW/SS]
by PS[M3/SS] phase, what is required by the Maxwell construction
(see Eq.~(\ref{Max}) and {\em Theorem 1}). With increasing $|U|$
the threshold and the extent of PS[CDW/SS] phase are decreased
and, for large enough $|U|$, the M3 phase is stabilized for all
$W>0$ at $n=1$, as shown in Fig.~(2b). The area of PS[CDW/SS]
disappears at all, in favor of mixed phase.

In both Figs~(2a) and (2b) there is also a line "cutting off" the
upper right-hand corner. This is the boundary for existence of
bound pairs of extended s-wave symmetry, given by
Eq.~(\ref{wcrmy}). In view of Randeria's notion\cite{randeria},
about necessity of such pairs for existence of superconductivity
of this symmetry it can be used as an additional bound for the
phase diagram\cite{bak2sol}. Above these lines no
superconductivity can survive so the NO phase enters diagram (to
be sure of that we must consult again the free energy diagrams).
Let's note, that it is only an approximation, as we did not
calculated extended s-wave superconductivity but only on-site,
pure s. Preliminary results show nevertheless that the parameter
of extended superconductivity $\Delta_\gamma$ is two orders of
magnitude smaller than $\Delta_0$ along the boundary PS/SS state.
Also $\eta$-pairing can have extended component, which will be
stabilized by $W>0$, due to negative sign at the extended part of
the pairing potential, unlike the case of extended s-wave
superconductivity with $Q=0$. This fact will still enhance the
main tendency outlined in this paper.

Inclusion of the normal Fock parameter into the calculations will
bring about the increase of the bandwidth, thus effectively
pushing the system into weak-coupling limit. It can cause the
increasing of thresholds but should not change qualitatively the
main results of this paper.

\section{Conclusions}
In the paper a phase diagram of the
extended Hubbard model with attractive on-site and repulsive, nn,
intersite interactions was analyzed in the Hartree-Fock
approximation. A possibility of stabilizing the mixed phase in the
model with short range, nn interactions only, was shown.
Stabilization requires the existence of the three orderings:
superconductivity, $\eta$-superconductivity and charge density
waves. The extent of the mixed phase increases with increasing
$|U|$.

The other result of the paper comes from the solution of the
Schrodinger equation for bound pairs of extended s-wave symmetry .
The area of the normal phase is introduced into the phase diagram
due to this criterion.

\subsection*{Acknowledgements}
I acknowledge discussions with R. Micnas, S. Robaszkiewicz, T.
Kostyrko and support from the Foundation for Polish Science. This
work was also supported by the Polish State Committee for
Scientific Research (KBN), Project No. 1 P03B 084 26.

\end{document}